\documentclass[10pt]{wlscirep}
\usepackage[utf8]{inputenc}
\usepackage[position=t,singlelinecheck=off]{subfig}
\usepackage[T1]{fontenc}
\usepackage{setspace}
\title{The Algebraic Expressions of Huygens Principle and Holographic Principle of Light}

\author[1,*]{Malong Fu}
\author[2,*]{Yang Zhao}
\affil[1]{National Key Laboratory of Science and Technology on Miro/Nano Fabrication, Shanghai Jiao Tong University, Shanghai 200030, People's Republic of China}
\affil[2]{Department of Electrical Engineering and Computer Science, Lassonde School of Engineering, York University, Toronto, ON M3J1P3 Canada}

\affil[*]{fumelon@sjtu.edu.cn}

\begin{abstract}
\textbf{Huygens principle (HP) is the cornerstone of wave optics, its mathematical model is a boundary value problem of wave equation. The solutions of this mathematical model should be $\partial u/\partial n$ independent and satisfy the form of retarded potential. In the engaged formulas, only the Rayleigh-Sommerfeld diffraction formula (RSDF) satisfies these two restrictions. Unfortunately, the HP requires spherical boundary, while the boundary of RSDF is an infinite plane. Besides that, we find the the geometric constructions of HP and holographic principle of light (HPL) are complementary. Here we derive out the complete expressions of HP and HPL with spherical boundary, based on the method of images. Furthermore, the HP, HPL and RSDF are combined into one new principle that if the boundary of a vacuum region is a spherical surface or an infinite plane, all the light in this vacuum region is determined by the light on the boundary.}
\end{abstract}
\begin{document}
\captionsetup[figure]{labelfont={bf},labelformat={default},labelsep=period,name={Fig.}}
\flushbottom
\maketitle
%
%
\thispagestyle{empty}

\section*{Introduction}

\begin{figure}[ht]
\centering
\includegraphics[width=.45\linewidth]{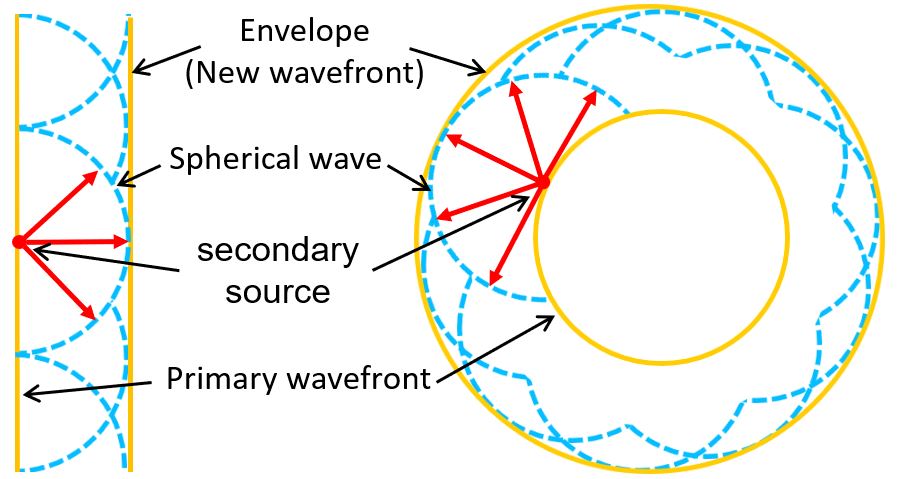}
\caption{\textbf{The geometric construction of the initial version of HP.}}
\label{fig:1}
\end{figure}

Huygens principle (HP) is a wave theory to explain the propagation of light in free space\cite{Arnold2004Intro}. With centuries of efforts in both physical and mathematical theory, many explanations are proposed to try to give a precise comprehension of HP. The initial version proposed by Huygens which is shown in Fig.\ref{fig:1}, is that every point on the primary wavefront can be regarded as a secondary source emitting spherical wave, the envelop of these spherical waves constitutes the new wavefront\cite{Max1997Foundations}. In other words, the primary wave front determines the subsequent process of light propagation. 

Based on it, Fresnel introduced the assumptions of the interference between spherical waves and the inclination factor. However, his formula is not strictly derived in mathematics. Thanks to the electromagnetic wave property of the light identified by Maxwell, researchers realized that the propagation of light can be taken as the wave equation in vacuum. Then, Kirchhoff proposed the first wave equation based expression of HP, i.e. the general form of Kirchhoff's theorem as given by Eq.(\ref{eq:1})\cite{Max1997Elements}.

\begin{align}
u(P_1,t)=\frac{1}{4\pi }\iint_{S_0}^{ }\left [ u(P_0,t-\frac{r_{01}}{v})\frac{\partial }{\partial n}\frac{1}{r_{01}}-\frac{1}{v r_{01}}\frac{\partial }{\partial n}\frac{\partial u(P_0,t-\frac{r_{01}}{v})}{\partial t} -\frac{1}{r_{01}}\frac{\partial u(P_0,t-\frac{r_{01}}{v})}{\partial n}\right ] {\rm d}S
\label{eq:1}
\end{align}

Where $u(P,t)$ represents the light disturbance at point $P$ and instant $t$, $P_1$ is the point to be calculated, $S_0$ stands for the primary wavefront, $P_0$ represents any point on $S_0$, $r_{01}$ denotes the distant between $P_0$ and $P_1$, $v$ is the light speed in vacuum, $n$ gives the normal of $S_0$. Unfortunately, it had been proved by Poincar$\acute{\rm e}$ and Sommerfeld that $\partial u/\partial n$ could cause a mathematical paradox\cite{J1996Historical,Sommerfeld1954Huygens}. Sommerfeld adopted the method of images to improve the Kirchhoff’s deduction\cite{Arnold2004Intro}. As a result, the general form of Rayleigh-Sommerfeld diffraction formula (RSDF) came out as in Eq.(\ref{eq:2})\cite{J1996Generalization}.
\begin{align}
u( P_{1},t  )=\iint_{S_0}^{ }\frac{\cos \alpha }{2\pi r_{01}}(\frac{1}{v}\frac{\partial }{\partial t}+\frac{1}{r_{01}}) u( P_{0},t-\frac{r_{01}}{v} ){\rm d}S \label{eq:2}
\end{align}

Where $\alpha $ is the angle between $n$ and $r_{01}$, the term $1/r_{01}$ is often omitted since $r_{01}$ is far greater than the wave length of light. The RSDF could be the most ideal expression of HP, but the integration surface $S_0$ must be an infinite plane, thereby the geometric construction of RSDF is just the left of Fig.\ref{fig:1}. However, the right of Fig.\ref{fig:1}indicates the HP should be also valid for spherical integration surface. 

On the other hand, it can be found from these expressions that the time variable $t$ always follows the form $t-1/r_{01}$. The physical meaning of $t-1/r_{01}$ is clear: the light at $P_1$ at instant $t$ is caused by the light at $P_0$ at instant $t_0$ after propagating a certain distance $r_{01}$, while due to the limited and invariant speed of light, the time difference between $t$ and $t_0$ is $1/r_{01}$. That is, the light produced at $P_0$ at other instant except $t-1/r_{01}$, gives no contribution to the light at $P_1$ at instant $t$. This effect is known as retarded potential in physics. Mathematically, for all solutions of wave equation, the time variable $t$ must follow the form $t\pm 1/r_{01}$, where $t+1/r_{01}$ represents the advanced potential, e.g., the general solution of the one-dimensional or the spherically symmetric wave equation\cite{R2004Infinite}. The advanced potential usually be discarded for conflicting both with experience and elementary notions of macroscopic causality\cite{J1949Classical}.

\begin{figure}[htbp]
\captionsetup[subfigure]{labelformat=simple}
\centering
\renewcommand{\thesubfigure}{\alph{subfigure}}
\subfloat[]{
\label{fig:subfig:1a}
\includegraphics[width=.23\linewidth]{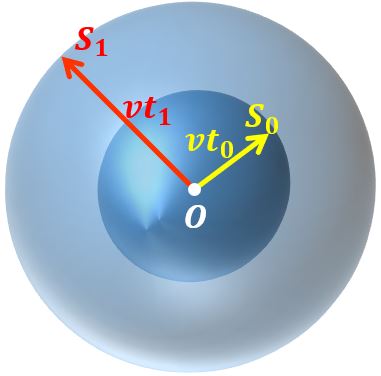}
}
\qquad\qquad
\subfloat[]{
\label{fig:subfig:1b}
\includegraphics[width=.23\linewidth]{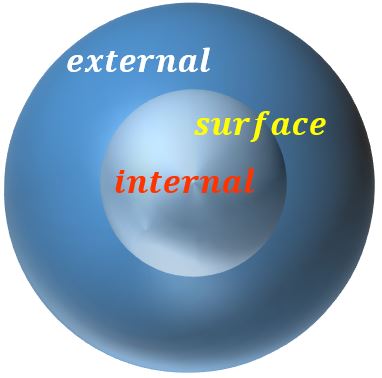}
}
\caption{\textbf{The geometric construction of HP and holographic principle.} (\textbf{a})The geometric construction of HP. (\textbf{b})The geometric construction of holographic principle.}
\label{fig:2}
\end{figure}

A more profound and realistic version of HP was proposed by J. Hadamard. It describes the propagation of a given spherical light in free space\cite{Jacques1923fundamental}. The geometric construction of J. Hadamard’s version is illustrated in Fig.\ref{fig:subfig:1a}. It shows that the light produced by point $O$ successively arrives surface $S_0$ at instant $t_0$, and arrives surface $S_1$ at instant $t_1$. He raised the physical essence of HP as that the effect of the light from $O$ to $S_1$, can be replaced by the effect from $S_0$ to $S_1$. 

The holographic principle manifests that all the information in a three-dimensional space is coding on the two-dimensional surface which wrapping this three-dimensional space\cite{G1993Dimensional,R2002holographic}. That is, the information on the surface of a region determines all the events inside the region. As a matter of logic, any effect from external to internal of the surface, can be replaced by the effect from the surface to internal which is shown in Fig.\ref{fig:subfig:1b}. Thus, the geometric constructions of HP and holographic principle are complementary in terms of the spherical surface (Fig.\ref{fig:2}). Therefore, we can safely give the holographic principle of light (HPL): the light at any point within a vacuum region which is surrounded by a spherical surface, is determined by the light on the spherical surface. The vacuum region in this paper means a region has no light source or any matter, the light in vacuum region is absolutely free to propagate and satisfies the wave equation. Correspondingly, the source region contains light source or matters, and the light in it doesn't always satisfy the wave equation.

It is not difficult to transform the geometric constructions of RSDF, HP and HPL to a same algebraic problem, which can be expressed as:
\begin{align}
\left\{\begin{matrix}
\triangledown ^2u-\frac{1}{v^2} \frac{\partial ^2 }{\partial t^2}u=0 \\ \\
 u(P_0,t)=g(t),\frac{\partial }{\partial t}u(P_0,t)=\dot{g}(t),P_0\in S_0
\end{matrix}\right. 
\label{eq:3}
\end{align}

Consequently, the mathematical model of RSDF, HP and HPL is the boundary value problem of wave equation, although the form of definite condition is Cauchy condition. Since these principles come from practical problems, the solution of the mathematical model must be existing, unique, $\partial u/\partial n$ independent and conforming to the retarded potential. Specially, when $S_0$ divides the whole space into source region and vacuum region two parts, three cases should be discussed: (1) If $S_0$ is an infinite plane, the solution is the RSDF; (2) If $S_0$ is a spherical surface wrapping the source region, the solution is the expression for HP; (3) If $S_0$ is a spherical surface wrapping the vacuum region, the solution is the expression for HPL.

The solutions of case (2) and (3) are given in this paper. According to the Fourier Transform (FT), the wave equation can be transformed to Helmholtz equation (HE). To solve the HE, two elegant Green’s functions (GF) are proposed respectively for HP and HPL by using the method of images. After obtaining the solutions for HE, the solutions of the wave equation are finally achieved via FT. The result shows that the nature of HP, HPL and RSDF are identical.

\section*{Solving the Mathematical Model of HP and HPL}
The object is to find the special formulas from wave equation to calculate $u$ at any point, base on the given $u$ and $\frac{\partial u}{\partial t}$ on spherical surface $S_0$. Considering any disturbance at a fixed point can be represented by the sum of infinite rotations on complex plane, the propagation issue of light disturbance could be transformed to the steady state issue of complex amplitude. According to the method of images, two innovative GFs are respectively proposed to solve the steady state issue for HP and HLP. Finally, the ideally expression pair is obtained via FT. 

\subsection*{FT and Steady State Issue} 
Monochrome disturbance of light is the simple harmonic wave in nature, it can be expressed as the following sine form at a fixed point $P$.
\begin{align}
F(P,f )\sin \left [ 2\pi f t+\phi  (P,f) \right ]  \label{eq:4}
\end{align}

Where $F$, $f $ and $\phi$ are amplitude, frequency and initial phase respectively. Bring Euler's formula into Eq.(\ref{eq:4}), apply odd extensions: $F(P,-f )=-F(P,f )$, $\phi (P,-f )=-\phi(P,f )$, and meanwhile introduce $U(P,f )=i F(P,f )\exp \left [ -i\phi(P,f )\right ]/2$. We can have:
\begin{align}
\begin{split}
F(P,f )\sin \left [ 2\pi f t+\phi (P,f) \right ]&=F(P,f)\cdot \frac{\exp \left [ i(2\pi f t+\phi) \right ]-\exp \left [ -i(2\pi f t+\phi)\right ]}{2i}\\
&=U(P,f )\exp (-i2\pi f t )+U(P,-f )\exp (i2\pi f t )  \label{eq:5}
\end{split}
\end{align}

Where $U(P,f )\exp ( -i2\pi f t)$ represents a clockwise rotation with angular velocity $2\pi f$, radius $F/2$ and initial phase $\pi/2-\phi$ on complex plane. $U$ stands for complex amplitude, which is different from the definition of Goodman\cite{J1996Some}. Thus the monochromatic disturbance can be represented by two rotations on complex plane, one is clockwise, the other is counterclockwise. Any disturbance at a fixed point can be regarded as the superposition of infinite monochromatic disturbance:
\begin{align}
\begin{split}
u(P,t )&=\int_{0 }^{\infty }F(P,f )\sin \left[ 2\pi f t+\phi  (P,f) \right ]  {\rm d}f\\
&=\int_{0 }^{\infty }\left [ U(P,f )\exp (-i2\pi f t )+U(P,-f )\exp (i2\pi f t ) \right ]{\rm d}f\\
&=\int_{-\infty }^{\infty }U(P,f )\exp ( -i2\pi f t) {\rm d}f\\
&=\frac{1}{2\pi }\int_{-\infty }^{\infty }U(P,\omega )\exp ( -i\omega t) {\rm d}\omega  \label{eq:6}
\end{split}
\end{align}

Where $\omega=2\pi f$ is the circular frequency. Eq.(\ref{eq:6}) is actually the FT, it means any disturbance can be decomposed into infinite rotations on complex plane. Actually, the $-i$ also can be replaced by $i$ with modifying the expression of $U$. The posteriori analysis indicates that the $-i$ is suitable for the retarded potential, while $i$ for the advanced potential. If each rotation of a disturbance satisfies the wave equation, the disturbance satisfies too. Therefore replace the light disturbance $u$ by $U(P,\omega )\exp ( -i\omega t)$ in wave equation:
\begin{align}
\triangledown ^2 U(P,\omega) +k^2 U(P,\omega)=0  \label{eq:7}
\end{align}

Eq.(\ref{eq:7}) is the famous HE, where $k$ is the wave number of the monochromatic light and $k=\omega /v$. By the way, the HE is an elliptic partial differential equation, which is actually in charge of the steady state issue, i.e., the $U(P,\omega)$ represents a complex amplitude field distributing throughout the whole space. 

If the relationship of complex amplitude of monochromatic disturbance between $P_0$ and $P_1$ is derived from HE, the expressions of HP and HPL can be deduced by FT, thus the goal is to solve the HE. 

\subsection*{GF and Method of Images} 
The analytical solution of HE is usually obtained through the Green's theorem: if $U$ and $G$ are both single-valued functions of position and twice continuously differentiable in a simply connected domain $V$ where is surrounded by surface $S$, the Green's second identity is set up as:
\begin{align}
\iiint_{V}^{ }(U \triangledown ^2 G -G\triangledown ^2 U ) {\rm d}V=\iint_{S}^{ }(U\frac{\partial G}{\partial n}-G\frac{\partial U}{\partial n}){\rm d}S
\label{eq:8}
\end{align}

Where $n$ stands for the outward normal of $S$, $G$ is the GF for the steady state issue, whose physical meaning is that the influence from one point of source to the point out of source\cite{R2004Green}. If $G=0$ on $S$, the term $G\partial u/\partial n$ can be ignored. According to the initial version of HP: every point on the wavefront can be regarded as a secondary source which emits spherical wave, thus the GF for HP should be as:
\begin{align}
G(P_1,P_0)=\frac{\exp (i\vec{k}\cdot \vec{r}_{01})}{r_{01}}=\frac{\exp (i k r_{01})}{r_{01}}
\label{eq:9}
\end{align}

Where $\vec{k}$ is the wave vector of monochromatic light, whose direction represents the propagation direction of light. $\vec{r}_{01}$ is the position vector from $P_0$ to $P_1$. Meanwhile according to the Maxwell's reciprocity of GF\cite{R2004Green}:
\begin{align}
G(P_0,P_1)=G(P_1,P_0)=\frac{\exp (i k r_{01})}{r_{01}}
\label{eq:10}
\end{align}

For making it appear at the final expression, Kirchhoff gives the GF as:
\begin{align}
G(P_x,P_1)=\frac{\exp (i\vec{k}\cdot \vec{r}_{x1})}{r_{x1}}=\frac{\exp (i k r_{x1})}{r_{x1}}
\label{eq:11}
\end{align}

Where $P_x$ is any point in the whole space except $P_1$, $\vec{r}_{x1}$ is the position vector from $P_x$ to $P_1$. Eq.(\ref{eq:11}) manifests that $P_1$ is a negative source which receives spherical wave. As mentioned in introduction, it will bring $\partial u/\partial n$. 

For any different two points in space, if the distances from each point of some points to the two points are in fixed proportion, the set of these points is a surface. In geometry, the two points are known as mirror points about the surface. Specially, if the proportion is equal to one, the surface is an infinite plane, and if not equal to one, the surface is a spherical surface (Apollonius spherical surface). Sommerfeld employs the former, and the latter will be adopted in this paper. 

To eliminate $\partial u/\partial n$, Sommerfeld adds an item on Kirchhoff's GF:
\begin{align}
G(P_x;P_1,P_2)
=\frac{\exp (i\vec{k}_1\cdot \vec{r}_{x1})}{r_{x1}}+\frac{\exp (i\vec{k}_2\cdot \vec{r}_{x2}+i\pi )}{r_{x2}}
=\frac{\exp (i k r_{x1})}{r_{x1}}-\frac{\exp (i k r_{x2})}{r_{x2}}
\label{eq:12}
\end{align}

Where $P_2$ is the mirror points of $P_1$ about $S_0$ which is an infinite plane, $\vec{r}_{x2}$ is the position vector from $P_x$ to $P_2$. If $P_x$ belongs to $S_0$, then $G=0$. Eq.(\ref{eq:12}) indicates $P_1$ and $P_2$ are two negative sources, thus $\vec{k_1}$ and $\vec{k_2}$ represent the wave vector of two converged spherical wave respectively and $ | \vec{k}_1  |=| \vec{k}_2  |=k$. Back to our case, a different GF should be found to remove the $\partial u/\partial n$ while $S_0$ is a spherical surface.

\subsection*{Proposed the Expression for HP} 
\begin{figure}[ht]
\centering
\includegraphics[width=.28\linewidth]{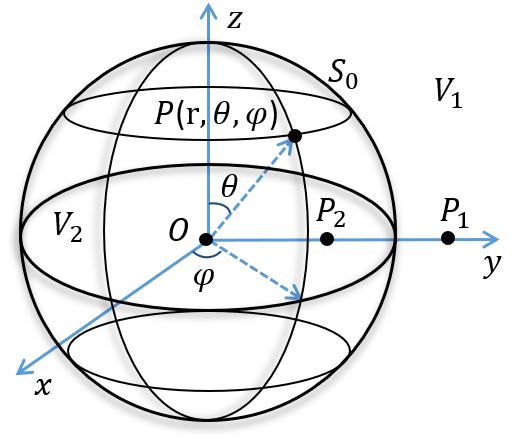}
\caption{\textbf{Polar coordinate system.} $r$: radial distance. $\theta$: elevation angle. $\varphi $: the azimuth angle}
\label{fig:3}
\end{figure}

Before apply the GF, a coordinate system is necessary. Firstly, assuming the centre and radius of $S_0$ are $O$ and $b$ respectively, and the distance from $P_1$ to $O$ is $c$. Secondly, put the $S_0$ in a polar coordinate system and make the coordinate of $P_1$ as $(c,\pi /2,\pi /2)$. Thirdly, note the region where $r>b$ as $V_1$ and the region where $r<b$ as $V_2$, thus for HP, $V_1$ is the vacuum region and $V_2$ the source region. Which is constructed in Fig.\ref{fig:3}.
For any point $P_1$, one can always find its mirror point $P_2(a,\pi /2,\pi /2)$ about $S_0$, and $a$, $b$, $c$ satisfy $0<a<b<c$ and $c=b^2/a$. According to the characters of the Apollonius spherical surface, meanwhile note the distance from $P_0$ to $P_2$ as $r_{02}$, it is tenable that:
\begin{align}
r_{01}=\frac{c}{b}r_{02}=\frac{b}{a}r_{02}
\label{eq:13}
\end{align}

The special GF for HP is given as: 
\begin{align}
G_1 ( P_{x};P_1,P_2 )=\frac{\exp( i\vec{k_1} \cdot \vec{r}_{x1}) }{r_{x1}}+\frac{b}{c}\frac{\exp ( i \vec{k_2} \cdot \vec{r}_{x2} +i\pi)}{r_{x2}}=\frac{\exp ( i k r_{x1}  )}{r_{x1}}-\frac{b}{c}\frac{\exp ( i \frac{c}{b}k r_{x2}  )}{r_{x2}}
\label{eq:14}
\end{align}

\begin{figure}[htbp]
\centering
\includegraphics[width=.35\linewidth]{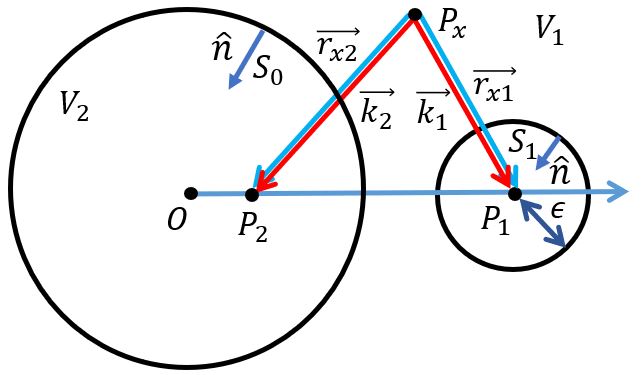}
\caption{\textbf{The illustration of $G_1$.} $S_1$: a spherical surface whose centre is $P_1$ and radius is $\varepsilon $. $S_0$ and $S_1$ are the boundary of $V_1$. $\hat{n}$: the outward normal of $V_1$'s boundary. The red lines represent the propagation direction of spherical wave.}
\label{fig:4}
\end{figure}

Where $ | \vec{k}_1  |=k$ and $| \vec{k}_2  |=ck/b$. $G_1$ is illustrated in Fig.\ref{fig:4}, the surface at infinity is excluding due to Sommerfeld radiation condition\cite{Sommerfeld1954Infinite}. Bring Eq.(\ref{eq:13}) into Eq.(\ref{eq:14}), $G_1$ is vanished on $S_0$, which just meets the above expect.

Bring Eq.(\ref{eq:7}) and Eq.(\ref{eq:14}) into the left of Eq.(\ref{eq:8}):
\begin{align}
\begin{split}
&\iiint_{V_1}^{ }(U \triangledown ^2 G_1 -G_1\triangledown ^2 U ) {\rm d}V\\
=&\iiint_{V_1}^{ }U\left [ -k^2 \frac{\exp ( i k r_{x1}  )}{r_{x1}}+(\frac{c}{b}k)^2\frac{b}{c}\frac{\exp ( i \frac{c}{b}k r_{x2}  )}{r_{x2}} \right ]+\left [ \frac{\exp ( i k r_{x1}  )}{r_{x1}}-\frac{b}{c}\frac{\exp ( i \frac{c}{b}k r_{x2}  )}{r_{x2}} \right ]k^2 U \\
=&(\frac{c^2-b^2}{b c})k^2\iiint_{V_1}^{ }U(P_x,\omega )\frac{\exp ( i \frac{c}{b}k r_{x2}  )}{r_{x2}} {\rm d}V
\end{split}
\label{eq:15}
\end{align}

The boundary of $V_1$ includes $S_0$ and $S_1$ two parts, as a result, while notice $G_1=0$ on $S_0$, the right of Eq.(\ref{eq:8}) becomes:
\begin{align}
\iint_{S_0}^{ }U\frac{\partial G_1}{\partial n}{\rm d}S+\iint_{S_1}^{ }(U\frac{\partial G_1}{\partial n}-G_1\frac{\partial U}{\partial n}){\rm d}S
\label{eq:16}
\end{align}

It can be easily confirmed that\cite{Sommerfeld1950Vector,Sommerfeld1954Huygens}:
\begin{align}
\lim_{\varepsilon \rightarrow 0}\iint_{S_1}^{ }(U\frac{\partial G_1}{\partial n}-G_1\frac{\partial U}{\partial n}){\rm d}S=-4\pi U(P_1, \omega)
\label{eq:17}
\end{align}

\begin{figure}[ht]
\centering
\includegraphics[width=.33\linewidth]{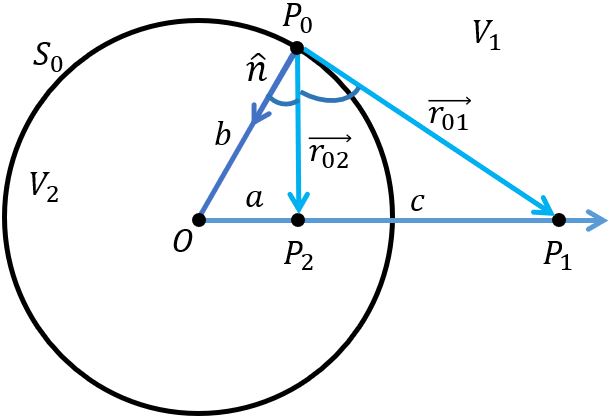}
\caption{\textbf{Surface integral on $S_0$.} $\vec{r}_{02}$: the position vector from $P_0$ to $P_2$.}
\label{fig:5}
\end{figure}

The surface integral on $S_0$ is Eq.(\ref{eq:18}) and illustrated in Fig.\ref{fig:5}.
\begin{align}
\begin{split}
\iint_{S_0}^{ }U\frac{\partial G_1}{\partial n}{\rm d}S&=\iint_{S_0}^{ }U(P_0, \omega)\left [ \frac{\hat{n} \cdot \vec{r}_{01}}{r_{01}}(i k-\frac{1}{r_{01}})\frac{\exp ( i k r_{01} )}{r_{01}}-\frac{\hat{n} \cdot \vec{r}_{02}}{r_{02}}(i \frac{c}{b}k-\frac{1}{r_{02}})\frac{b}{c}\frac{\exp ( i \frac{c}{b}k r_{02} )}{r_{02}} \right ]{\rm d}S\\
&=\iint_{S_0}^{ }(\frac{\hat{n} \cdot \vec{r}_{01}}{r_{01}}-\frac{c}{b}\frac{\hat{n} \cdot \vec{r}_{02}}{r_{02}})(i k-\frac{1}{r_{01}})U(P_0, \omega)\frac{\exp (i k r_{01})}{r_{01}}{\rm d}S\\
&=\iint_{S_0}^{ }\frac{b^2-c^2}{b r_{01}}(i k-\frac{1}{r_{01}})U(P_0, \omega)\frac{\exp (i k r_{01})}{r_{01}}{\rm d}S
\label{eq:18}
\end{split}
\end{align}

Where $\hat{n} \cdot \vec{r}_{01}/r_{01}=\cos \angle O P_0 P_1$ and $\hat{n} \cdot \vec{r}_{02}/r_{02}=\cos \angle O P_0 P_2$, they both can be calculated by the cosine theorem. Bring Eq.(\ref{eq:15}), Eq.(\ref{eq:17}) and Eq.(\ref{eq:18}) into Eq.(\ref{eq:8}), the complex amplitude of $P_1$ can be expressed as:
\begin{align}
U(P_1,\omega )=\frac{1}{4\pi }\iint_{S_0}^{ }\frac{b^2-c^2}{b r_{01}}(i k-\frac{1}{r_{01}})U(P_0, \omega)\frac{\exp (i k r_{01})}{r_{01}}{\rm d}S+\frac{1}{4\pi }(\frac{b^2-c^2}{b c})k^2\iiint_{V_1}^{ }U(P_x,\omega )\frac{\exp ( i \frac{c}{b}k r_{x2}  )}{r_{x2}} {\rm d}V
\label{eq:19}
\end{align}

Eq.(\ref{eq:19}) is the solution of the HE, thus it is a formula for steady state issue. The formula for propagation issue can be obtained via FT. Bring Eq.(\ref{eq:19}) into Eq.(\ref{eq:6}):
\begin{align}
\begin{split}
u ( P_{1},t  )&=\frac{1}{2\pi }\int_{-\infty }^{\infty }\left [ \iint_{S_0}^{ }\frac{b^2-c^2}{4\pi b r_{01}}(i k-\frac{1}{r_{01}})U(P_0, \omega)\frac{\exp (i k r_{01})}{r_{01}}{\rm d}S+\frac{(b^2-c^2)k^2}{4\pi b c}\iiint_{V_1}^{ }U(P_x,\omega )\frac{\exp ( i \frac{c}{b}k r_{x2}  )}{r_{x2}} {\rm d}V \right ]\exp ( -i\omega t) {\rm d}\omega\\
&=\frac{1}{4\pi }\iint_{S_0}^{ }\frac{b^2-c^2}{b r_{01}^2}\left \{ \frac{1}{2\pi }\int_{-\infty }^{\infty }(\frac{i\omega }{v}-\frac{1}{r_{01}}) U(P_0,\omega)\exp \left [ -i\omega( t-\frac{r_{01}}{v} )\right] {\rm d}\omega \right \} {\rm d}S\\
&+\frac{1}{4\pi }(\frac{b^2-c^2}{b c})\iiint_{V_1}^{ }\left \{ \frac{1}{2\pi }\int_{-\infty }^{\infty }\frac{\omega ^2}{v^2r_{x2}}U(P_x,\omega )\exp \left [ -i\omega( t-\frac{\frac{c}{b}r_{x2}}{v} )\right ] {\rm d}\omega \right \} {\rm d}V\\
&=\frac{1}{4\pi }\iint_{S_0}^{ }\frac{c^2-b^2}{b r_{01}^2}(\frac{1}{v}\frac{\partial }{\partial t}+\frac{1}{r_{01}}) u ( P_{0},t-\frac{r_{01}}{v} ){\rm d}S
+\frac{1}{4\pi }\iiint_{V_1}^{ }(\frac{c^2-b^2}{b^2 v^2 \frac{c}{b} r_{x2}})\frac{\partial^2 }{\partial t^2}u(P_x,t-\frac{\frac{c}{b}r_{x2}}{v} ){\rm d}V
\label{eq:20}
\end{split}
\end{align}

The first term of right of Eq.(\ref{eq:20}) represents the contribution of the light at $P_0$ at instant $t-r_{01}/v$ to the light at $P_1$ at instant $t$, which conforms to the retarded potential. The last term represents the contribution of the light at $P_x$ at instant $t-c r_{x2}/b v$ to the light at $P_1$ at instant $t$. However as a matter of fact, only the light at $P_x$ at instant $t-r_{x1}/v$ can influence the light at $P_1$ at instant $t$. Meanwhile notice $c r_{x2}/b\neq r_{x1} $ except the situation that $P_x$ belong to $S_0$. Consequently, the last term should be zero in physics. Mathematically, all solutions of wave equation must satisfy the form of retarded potential or advanced potential; Eq.(\ref{eq:20}) is a solution of wave equation; And only if the last term is zero, Eq.(\ref{eq:20}) satisfies the form of retarded potential. Overall, the last term must be zero, therefore:
\begin{align}
\begin{split}
u ( P_{1},t  )=\frac{1}{4\pi }\iint_{S_0}^{ }\frac{c^2-b^2}{b r_{01}^2}(\frac{1}{v}\frac{\partial }{\partial t}+\frac{1}{r_{01}}) u ( P_{0},t-\frac{r_{01}}{v} ){\rm d}S
\label{eq:21}
\end{split}
\end{align}

Eq.(\ref{eq:21}) manifests that if the disturbance at all points on a spherical surface at all instant is known, the disturbance, caused by the light source inside the spherical surface, at any point out of the spherical surface at any instant can be calculated. That is exactly the algebraic expression of HP.

\subsection*{Proposed the Expressions for HPL} 
\begin{figure}[htpt]
\centering
\includegraphics[width=.35\linewidth]{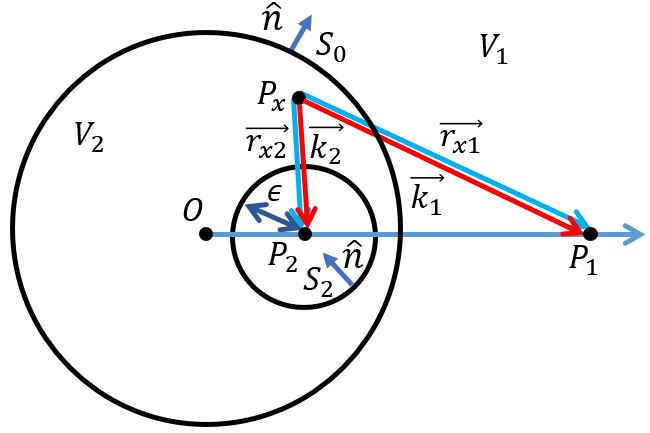}
\caption{\textbf{The illustration of $G_2$.} $S_2$: a spherical surface whose centre is $P_2$ and radius is $\varepsilon $. $S_0$ and $S_2$ are the boundary of $V_2$. $\hat{n}$: the outward normal of $V_2$'s boundary.}
\label{fig:6}
\end{figure}

For HPL, the source region is $V_1$ and the vacuum region is $V_2$. If consider $P_2$ as the point to be calculated, the $u(P_2,t)$ can be derived in the same way. In this situation, the coordinate system is still the same, but the GF changes to Eq.(\ref{eq:22}) which is shown in Fig.\ref{fig:6}. 
\begin{align}
\begin{split}
G_2( P_{x};P_2,P_1)
=\frac{\exp ( i \vec{k_2} \cdot \vec{r}_{x2} )}{r_{x2}}+\frac{b}{a}\frac{\exp ( i \vec{k_1} \cdot \vec{r}_{x1} +i\pi)}{r_{x1}}
=\frac{\exp ( i k r_{x2}  )}{r_{x2}}-\frac{b}{a}\frac{\exp ( i \frac{a}{b}k r_{x1}  )}{r_{x1}}
\label{eq:22}
\end{split}
\end{align}

Where $ | \vec{k}_2  |=k$ and $| \vec{k}_1 |=a k/b$, and $G_2=0$ on $S_0$. Employ exactly the same steps, the algebraic expression of HPL is obtained as:
\begin{align}
u ( P_2,t  )=\frac{1}{4\pi }\iint_{S_0}^{ }\frac{b^2-a^2}{b r_{02}^2}(\frac{1}{v}\frac{\partial }{\partial t}+\frac{1}{r_{02}}) u ( P_0,t-\frac{r_{01}}{v} ){\rm d}S
\label{eq:23}
\end{align}

Eq.~(\ref{eq:23}) manifests that if the disturbance at all points on a spherical surface at all instant is known, the disturbance, caused by the light source out of the spherical surface, at any point inside the spherical surface at any instant can be calculated.

\section*{Conclusions}
\begin{figure}[hbtp]
\captionsetup[subfigure]{labelformat=simple}
\centering
\renewcommand{\thesubfigure}{\alph{subfigure}}
\subfloat[]{
\label{fig:subfig:7a}
\includegraphics[width=.28\linewidth]{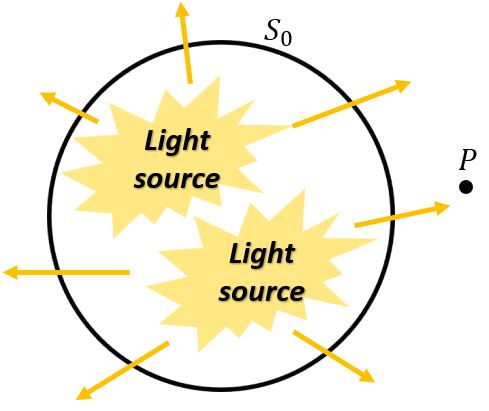}
}
\subfloat[]{
\label{fig:subfig:7b}
\includegraphics[width=.3\linewidth]{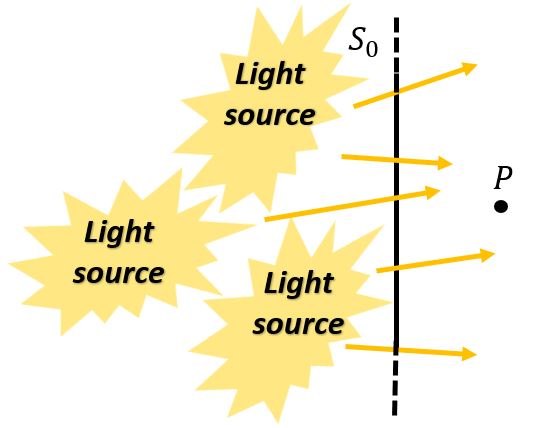}
}
\subfloat[]{
\label{fig:subfig:7c}
\includegraphics[width=.26\linewidth]{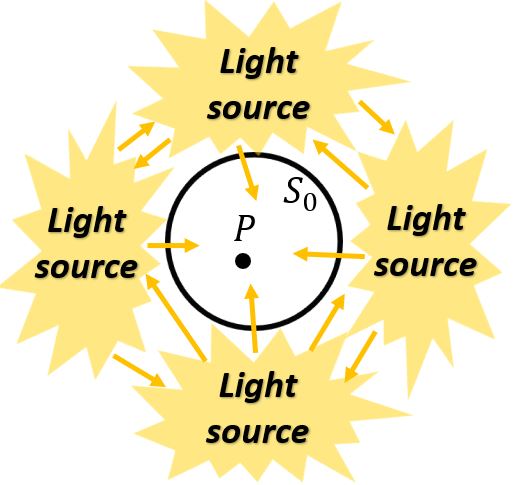}
}
\caption{\textbf{The pictures of HP, RSDF and HPL.} (\textbf{a})HP. (\textbf{b})RSDF. (\textbf{c})HPL. The light at point $P$ is determined by the light on $S_0$. From (\textbf{a}) to (\textbf{b}), the spherical surface $S_0$ which wrapping the light source expands into an infinite plane. From (\textbf{b}) to (\textbf{c}), the infinite plane $S_0$ rolls over to wrap $P$ and becomes a spherical surface again.}
\label{fig:7}
\end{figure}

The expression pair of HP and HPL have been derived in this paper, i.e., Eq.(\ref{eq:21}) and Eq.(\ref{eq:23}) respectively. There are three special issues should be emphasized. First, due to $a>0$ and $c=b^2/a$, the centre of $S_0$ and infinite $\infty $ should both be regarded as singularities, resulting the asymptotic values for the light disturbance at these points. Second, the initial version of HP claims $S_0$ is a wavefront, which implies the phase of light on $S_0$ must be identical and the direction of light is perpendicular to $S_0$. However during the deduction process, the two restrictions are not demanded, so no links between $S_0$ and wavefront. Third, the light source of Hadamard's version is at the centre of $S_0$, which is also not mentioned. Thus, the pictures of HP, RSDF and HPL can be illustrated in Fig.\ref{fig:7}. 

It is easy to discover from Fig.\ref{fig:7} that the nature of the three pictures are the same, except the different radius of $S_0$, e.g., if $b\rightarrow \infty$, $S_0$ will become an infinite plane, while Eq.(\ref{eq:21}) and Eq.(\ref{eq:23}) become Eq.(\ref{eq:2}). Therefore, they can be united as one principle: if the boundary of a vacuum region is a spherical surface or an infinite plane, the light in this vacuum region determined by the light on the boundary. Mathematically, for any physical disturbance, which satisfies the wave equation in a media region, these formulas and this principle should be also valid.

\section*{Acknowledgements}
M.F. acknowledges supports from the National Natural Science Foundation of China under Grant No. 60906053, 61204069, 61274118, 61306144 and 11605112 respectively.

\section*{Author contributions}
M.F. conducted the mathematical derivation, theory analysis and manuscript writing. Y.Z. conducted the manuscript writing.

\bibliographystyle{unsrt}

\bibliography{sample}

\begin{thebibliography}{10}
\urlstyle{rm}
\expandafter\ifx\csname url\endcsname\relax
  \def\url#1{\texttt{#1}}\fi
\expandafter\ifx\csname urlprefix\endcsname\relax\def\urlprefix{URL }\fi
\expandafter\ifx\csname doiprefix\endcsname\relax\def\doiprefix{DOI: }\fi
\providecommand{\bibinfo}[2]{#2}
\providecommand{\eprint}[2][]{\url{#2}}

\bibitem{Arnold2004Intro}
\bibinfo{author}{{A. Sommerfeld}}.
\newblock \bibinfo{title}{{Introduction}}.
\newblock In \emph{\bibinfo{booktitle}{Mathematical theory of diffraction}},
  \bibinfo{pages}{3--6},
  \doiprefix\url{https://doi.org/10.1007/978-0-8176-8196-8}
  (\bibinfo{publisher}{Springer Science+Bussiness Media},
  \bibinfo{year}{2004}).

\bibitem{Max1997Foundations}
\bibinfo{author}{{M. Born}} \& \bibinfo{author}{{E. Wolf}}.
\newblock \bibinfo{title}{{Foundations of geometrical optics}}.
\newblock In \emph{\bibinfo{booktitle}{Principles of optics: electromagnetic
  theory of propagation, interference and diffraction of light}},
  \bibinfo{pages}{132} (\bibinfo{publisher}{Pergamon}, \bibinfo{year}{1980}),
  \bibinfo{edition}{6th} edn.

\bibitem{Max1997Elements}
\bibinfo{author}{{M. Born}} \& \bibinfo{author}{{E. Wolf}}.
\newblock \bibinfo{title}{{Kirchhoff's diffraction theory}}.
\newblock In \emph{\bibinfo{booktitle}{Principles of optics: electromagnetic
  theory of propagation, interference and diffraction of light,6th ed}},
  \bibinfo{pages}{375--386} (\bibinfo{publisher}{Pergamon},
  \bibinfo{year}{1980}), \bibinfo{edition}{6th} edn.

\bibitem{J1996Historical}
\bibinfo{author}{{J. W. Goodman}}.
\newblock \bibinfo{title}{{Historical introduction}}.
\newblock In \emph{\bibinfo{booktitle}{Introduction to Fourier optics}},
  \bibinfo{pages}{32--36} (\bibinfo{publisher}{McGraw-Hill},
  \bibinfo{year}{1996}), \bibinfo{edition}{2th} edn.

\bibitem{Sommerfeld1954Huygens}
\bibinfo{author}{{A. Sommerfeld}}.
\newblock \bibinfo{title}{{Huygens’ principle}}.
\newblock In \emph{\bibinfo{booktitle}{Optics, lecture on theoretical
  physics}}, \bibinfo{pages}{195--206} (\bibinfo{publisher}{Academic},
  \bibinfo{year}{1954}).

\bibitem{J1996Generalization}
\bibinfo{author}{{J. W. Goodman}}.
\newblock \bibinfo{title}{{Generalization to nonmonchromatic waves}}.
\newblock In \emph{\bibinfo{booktitle}{Introduction to Fourier optics}},
  \bibinfo{pages}{53--54} (\bibinfo{publisher}{McGraw-Hill},
  \bibinfo{year}{1996}), \bibinfo{edition}{2th} edn.

\bibitem{R2004Infinite}
\bibinfo{author}{{R. Haberman}}.
\newblock \bibinfo{title}{{Infinite space Green's function for the
  three-dimensional wave equation(Huygens' principle)}}.
\newblock In \emph{\bibinfo{booktitle}{Applied partial differential equations:
  with Fourier series and boundary value problems}}, \bibinfo{pages}{518--520}
  (\bibinfo{publisher}{Pearson Education}, \bibinfo{year}{2004}),
  \bibinfo{edition}{4th} edn.

\bibitem{J1949Classical}
\bibinfo{author}{{J. A. Wheeler}} \& \bibinfo{author}{{R. P. Feynman}}.
\newblock \bibinfo{journal}{\bibinfo{title}{Classical electrodynamics in terms
  of direct interparticle action}}.
\newblock {\emph{\JournalTitle{Rev. Mod. Phys.}}}
  \textbf{\bibinfo{volume}{21}}, \bibinfo{pages}{425},
  \doiprefix\url{https://doi.org/10.1103/RevModPhys.21.425}
  (\bibinfo{year}{1949}).

\bibitem{Jacques1923fundamental}
\bibinfo{author}{{J. Hadamard}}.
\newblock \bibinfo{title}{{The fundamental formula and the elementary
  solution}}.
\newblock In \emph{\bibinfo{booktitle}{Lectures on Cauchy's problem in linear
  partial differential equations}}, \bibinfo{pages}{53--57}
  (\bibinfo{publisher}{Yale University}, \bibinfo{year}{1923}).

\bibitem{G1993Dimensional}
\bibinfo{author}{{G. ’t Hooft}}.
\newblock \bibinfo{journal}{\bibinfo{title}{Dimensional reduction in quantum
  gravity}}.
\newblock {\emph{\JournalTitle{Salamfest}}} \bibinfo{pages}{284}
  (\bibinfo{year}{1993}).
\newblock \eprint{arXiv:gr-qc/9310026}.

\bibitem{R2002holographic}
\bibinfo{author}{{R. Bousso}}.
\newblock \bibinfo{journal}{\bibinfo{title}{The holographic principle}}.
\newblock {\emph{\JournalTitle{Rev. Mod. Phys.}}}
  \textbf{\bibinfo{volume}{74}}, \bibinfo{pages}{825},
  \doiprefix\url{https://doi.org/10.1103/RevModPhys.74.825}
  (\bibinfo{year}{2002}).

\bibitem{J1996Some}
\bibinfo{author}{{J. W. Goodman}}.
\newblock \bibinfo{title}{{Some mathematical preliminaries}}.
\newblock In \emph{\bibinfo{booktitle}{Introduction to Fourier optics}},
  \bibinfo{pages}{38--42} (\bibinfo{publisher}{McGraw-Hill},
  \bibinfo{year}{1996}), \bibinfo{edition}{2th} edn.

\bibitem{R2004Green}
\bibinfo{author}{{R. Haberman}}.
\newblock \bibinfo{title}{{Green's function for boundary value problems for
  ordinary differential equations}}.
\newblock In \emph{\bibinfo{booktitle}{Applied partial differential equations:
  with Fourier series and boundary value problems}}, \bibinfo{pages}{385--399}
  (\bibinfo{publisher}{Pearson Education}, \bibinfo{year}{2004}),
  \bibinfo{edition}{4th} edn.

\bibitem{Sommerfeld1954Infinite}
\bibinfo{author}{{A. Sommerfeld}}.
\newblock \bibinfo{title}{{Infinite domains and continuous spectra of eigen
  values. The condition of radiation}}.
\newblock In \emph{\bibinfo{booktitle}{Partial differential equations in
  physics}}, \bibinfo{pages}{188--200} (\bibinfo{publisher}{Academic},
  \bibinfo{year}{1949}).

\bibitem{Sommerfeld1950Vector}
\bibinfo{author}{{A. Sommerfeld}}.
\newblock \bibinfo{title}{{A fundamental theorem of vector analysis}}.
\newblock In \emph{\bibinfo{booktitle}{Mechanics of deformable bodies, lecture
  on theoretical physics}}, \bibinfo{pages}{36--40}
  (\bibinfo{publisher}{Academic}, \bibinfo{year}{1950}).

\end{thebibliography}

\end{document}